\documentclass[12pt]{article}

\usepackage{rangecite}
\usepackage{graphicx}

\def\be{\begin{equation}}
\def\ee{\end{equation}}
\def\bear{\begin{eqnarray}}
\def\eear{\end{eqnarray}}
\def\nn{\nonumber}

\def\half{{{1\over 2}}}

\begin{document}

%%%%%%%%%%%%%%%%%%%%%%%%%%%%%%%%%%%%%%%%%%%%%%%%%%%%%%%%%%%%%%%%%%%%
%  Title page                                                      %
%%%%%%%%%%%%%%%%%%%%%%%%%%%%%%%%%%%%%%%%%%%%%%%%%%%%%%%%%%%%%%%%%%%%

\begin{titlepage}

\flushright{arXiv:0711.2680v2~[hep-th]}

\vskip 1.2in

\begin{center}
{\Large{Quantum behaviour near a spacelike boundary\\
in the c=1 matrix model.}}
\vskip 0.5in
{Joanna L. Karczmarek}
\vskip 0.3in
{\it 
Department of Physics and Astronomy\\
University of British Columbia,
Vancouver, Canada}
\end{center}
\vskip 0.5in

\begin{abstract}
Certain time dependent configurations in the c=1 matrix model 
correspond to string theory backgrounds which have spacelike boundaries
and appear geodesically incomplete.  
We investigate  quantum mechanical properties of a class of such
configurations in the matrix model, 
in terms of  fermionic eigenvalues.  
We describe Hamiltonian evolution of the eigenvalue density
using several different time variables, some of which are infinite and
some of which are finite in extent.  We derive unitary transformations
relating these different descriptions, and
use those to calculate fermion correlators in
the time dependent background.
Using the chiral formalism,  we write
the time dependent configurations as
a state in the original matrix model Hilbert space. 
\end{abstract}

\end{titlepage}

%%%%%%%%%%%%%%%%%%%%%%%%%%%%%%%%%%%%%%%%%%%%%%%%%%%%%%%%%%%%%%%%%%%%
%  Begin here                                                      %
%%%%%%%%%%%%%%%%%%%%%%%%%%%%%%%%%%%%%%%%%%%%%%%%%%%%%%%%%%%%%%%%%%%%

\tableofcontents

\section{Introduction}

It is well known that the $c=1$ matrix model is equivalent to two dimensional
Liouville string theory\footnote{Good review articles 
include \cite{Klebanov:1991qa}, \cite{Ginsparg:1993is}, and
\cite{Polchinski:1994mb} (chapter 5).  The approach
to effective action of the matrix model used here 
is reviewed in chapter 3 of \cite{Alexandrov:2003ut}
and in the references therein, as well as in
\cite{Das:1992dm} and \cite{Jevicki:1993qn}.}.
This equivalence is an example of open-closed duality:
 the density of matrix eigenvalues (representing the tachyonic
mode of open strings attached to D0-branes) is directly related to the
closed string `tachyon' field.  Since the $c=1$ matrix model is solvable,
it provides us with an exact quantum mechanical solution
to string theory in two dimensions.

This framework is a nice toy model for the study of time dependent
backgrounds in string theory.
Time dependent solutions can easily be constructed
in the matrix model, as
there are no conceptual difficulties associated with time dependence
in quantum mechanics.
These solutions correspond to particular
time dependent string theory backgrounds. Correlators of
small fluctuations can be studied  in the matrix model
to probe the spacetime structure of string backgrounds.
Any conceptual difficulties associated with presence of
time dependence in string theory
can be resolved by going back to the unambiguous description
in terms of matrix quantum mechanics.

One of the essential features of the matrix model solution to 
Liouville string 
is that space ({\it{e.i}}., the Liouville direction) 
is emergent: it is constructed from the
collective motion of matrix eigenvalues.
Time dependent backgrounds for string theory are constructed in
the matrix model by considering large deviations from the 
static eigenvalue distribution.
An outstanding issue of this approach has been that
these deviations might be too large to live in the 
Hilbert space of the original matrix quantum mechanics,
which would complicate their interpretation.  We address this issue here.

For static Liouville backgrounds, the time variable in string theory 
is inherited from matrix quantum mechanics;
in time dependent solutions the original
quantum mechanical time is mixed with the emergent space dimension.
The emergent nature of space and the mixing with the time dimension
make these models particularly interesting, potentially leading to
insights into the question of emergent {\it time} in string theory.

In \cite{Karczmarek:2003pv}, 
certain time dependent solutions in the $c=1$ matrix model
were proposed, presenting a variety of physical scenarios
which were further studied in
\cite{Karczmarek:2004ph,Das:2004hw,Mukhopadhyay:2004ff,Karczmarek:2004yc}.  
Some of the most promising scenarios
correspond to spacetimes with
spacelike boundaries ${\cal I}^+$ and/or ${\cal I}^-$
\cite{Das:2004aq,Ernebjerg:2004ut,Das:2007vfb}.
The appearance of spacelike ${\cal I}^\pm$
is associated with the existence of cosmological horizons, and
is reminiscent of de Sitter spacetimes.
Some properties of such solutions were studied in 
\cite{Das:2004aq,Das:2007vfb}, 
from the point of view of the classical effective theory.
In the present paper, a foray is made into  quantum mechanical
description of those solutions. 

Here we explore, at the full  quantum level,
the relationship between different solutions of 
matrix quantum mechanics.  
One of the results of this paper is that
our time-dependent solutions do live in the same Hilbert 
space as the static ones, and therefore should be 
thought of as fluctuations in the original theory,
a point which has not been made clear before.

The main thrust of the paper is that
the same quantum mechanical evolution can be
described as taking either a finite or an infinite amount of
time, depending on the choice of the time variable. 
The appearance of a finite time variable 
is what leads to a spacelike future
boundary ${\cal I}^+$ in string theory.

The existence of these drastically different and yet equivalent descriptions 
is interesting in its own right.
It is often stressed that one of the difficulties with quantum gravity
is that, while quantum mechanics assumes 
the presence of an {\it{a priori}} time, general relativity 
has no preferred time direction.  Our simple example 
illustrates that the requirement of an a priori time
in quantum mechanics might not be as rigid a constraint as
it is thought to be.  Here, quantum mechanical evolution
is written in terms of one of two different time variables, which
have different topologies:
one is infinite in extent, and the other only semi-infinite.
This behaviour is quite generic in quantum mechanics;
here we simply find a specific instance of it which gives us insight into
the quantum mechanical properties of particular time dependent
solutions of the c=1 matrix model.

Our quantum correspondence 
between different solutions allows us to relate the exact
quantum correlators in the time dependent solution to those in the static
solution.
While beyond the scope of this paper, further exploration of 
quantum correlators near the boundary might
lead to a calculation of the conformal factor
of the spacetime metric (which is not computable from classical
information), and eventually shine light on the nature
of spacelike singularities in string theory.

The paper is organized as follows.  In the next section,
we briefly review the solutions of interest from \cite{Das:2004aq}, and
introduce some useful notation.  In section 3, we write down
the correspondence between different solutions in quantum mechanical
language, and explain why they all live in the same Hilbert space. 
In section 4, using the chiral formalism,we 
write down explicit linear transformations 
between the wavefunctions describing different solutions.
In section 5, we study the fermion correlators, 
and compare our results to predictions from 
classical collective field theory.  Finally, 
in section 6, we discuss a few interesting consequences
and suggest possible extensions of our work.  
Our discussion is limited to
matrix quantum mechanics side of the duality, except
for some comments in the last section.  A variety of useful formul\ae\
is collected in the Appendix.

\section{Time dependent solutions in classical effective theory}

The $c=1$ matrix model quantum mechanics has as its fundamental
degrees of freedom non-interacting fermions in upside down 
harmonic oscillator potential, with the Hamiltonian
\be
H = \half p^2 - \half x^2~.
\label{hamiltonian}
\ee
The curvature at the top of the potential is fixed by taking $\alpha'=1$ in
the corresponding Liouville string theory.
The effective (or bosonized) picture for this system is that of
a Fermi fluid moving in phase space $(x,p)$.  
Its dynamics can be described in terms of the 
density of this fluid.  In the classical limit, the
density takes on values of either 1 or 0, since the
Fermi fluid is incompressible.  Therefore, it is sufficient 
to specify the region where eigenvalues are
present, which is the Fermi sea in  phase space, bounded by a Fermi surface.
In the simplest case, this surface can be 
presented as its upper and lower branches at each
point $x$,  $p_\pm(x,t)$.
The local density of fermions in $x$-space 
is then given by the distance between the
two branches of $p$:
\be
\varphi(x,t) \equiv {1\over 2} \left ( p_+(x,t) - p_-(x,t)\right)~. 
\ee
Static Fermi surfaces are hyperbol\ae\ given by the  equation
\be
\label{static}
x^2 - p^2 = 2\mu~,~ \mathrm{or}~~
\varphi_0 = \sqrt{x^2 - 2\mu}~.
\ee
Any small fluctuation around this static background moves
along one branch of the hyperbola from $x=\infty$ towards
finite $x$ and back out to $x=\infty$ along the other branch.
This is captured by the effective action for small
fluctuations about the static solution which is given by
\bear
\label{action}
S = \int d\tau d\sigma & &\left\{ {1\over 2}
((\partial_\tau\eta)^2 - (\partial_\sigma \eta)^2) - { \sqrt{\pi}
\over 6 \varphi_0^2} (3(\partial_\tau\eta)^2(\partial_\sigma
\eta)+ (\partial_\sigma \eta)^3) \right. \nonumber \\ &
&\qquad\left. ~+~ \frac{(\partial_\tau \eta)^2}{2}\sum_{n=2}^{\infty}
 \left(-{\sqrt{\pi} (\partial_\sigma \eta) \over \varphi_0^2}\right)^n
\right\}~. 
\eear
$\eta$ here is the fluctuation of fermion density about
its static configuration
\be
\varphi(x,t) = \varphi_0(x) + \sqrt \pi ~ \partial_x \eta(x,t)~,
\ee
and the coordinates $\sigma$ and $\tau$ are related to $x$ and $t$
via
\be
x = \sqrt {2 \mu} \cosh \sigma ~,~~~~~~~~~~~~t = \tau~.
\ee
Note that $\sigma$ is defined on the interval $[0,\infty)$ and there is
a Dirichlet boundary condition for $\eta$ at $\sigma = 0$.
Effectively, the fluctuations live on patch in $(\sigma, \tau)$
space pictured in a Penrose diagram in figure 1(a).  
The $\sigma$ and $\tau$ variables are related to the
spacetime coordinates in Liouville string theory by a nonlocal
transformation, whose exact form 
is known only at null asymptotia.
The characteristic length scale of the 
nonlocality is string length, and therefore
the Penrose diagram in figure 1(a), while strictly correct for the 
fermionic excitations of the matrix model, gives a good representation
of the spacetime of Liouville string on lengthscales longer than string 
lenght.

\begin{figure}[h]

\includegraphics[scale=0.8]{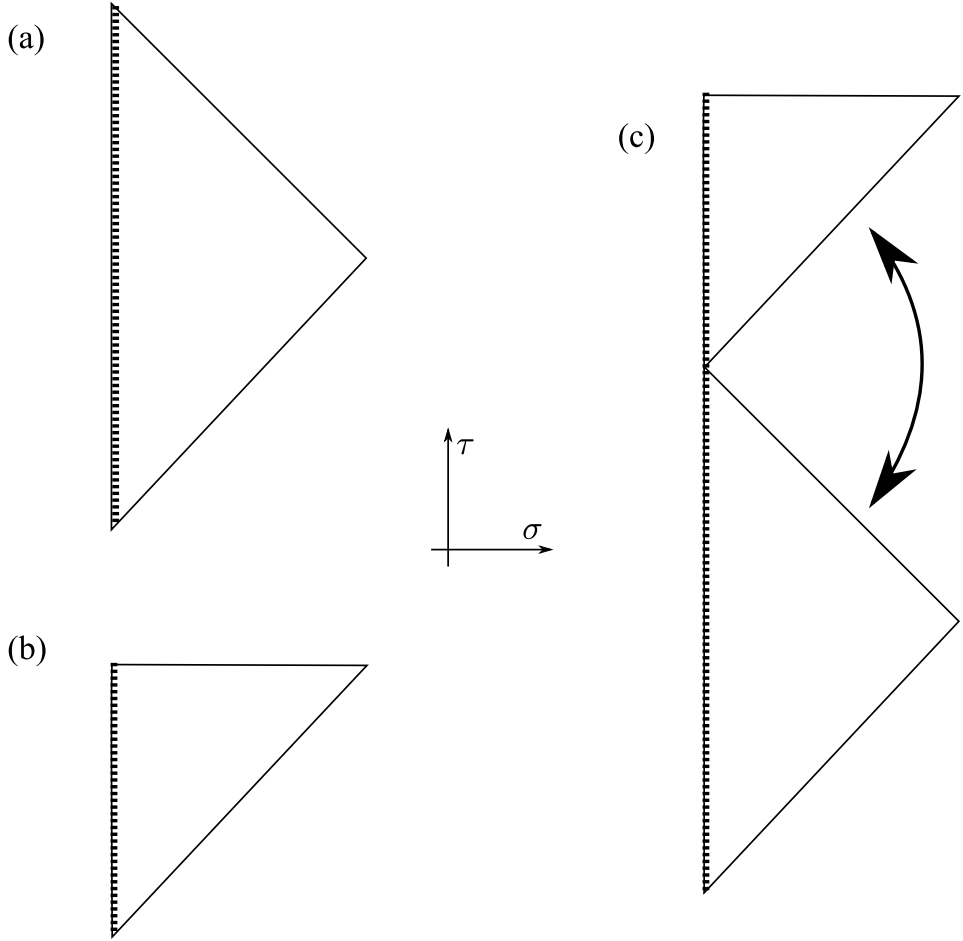}

\caption{The Penrose diagram of the causal structure of:
(a) static Liouville string; 
(b) spacetime resulting
from the closing hyperbola solution (\ref{closing});
(c) spacetime resulting
from the opening hyperbola solution (\ref{opening}).
In all three cases, the dashed line on the left hand side represents the
Liouville wall.}

\end{figure}

Having reviewed these basic facts, 
we will now start with simple time dependent solutions from 
\cite{Das:2004aq}
which exhibit a spacelike $\cal{I}^+$ and a null $\cal{I}^-$.

At the classical level, the first solution is given as a moving
Fermi surface
\be
\label{closing}
\left (x+p-e^{2t}(x-p)\right)(x-p) ~=~ 2\mu
\ee
Geometrically, this is a hyperbola which closes on itself
(see figure 2(b)).
Surprisingly, the following change of variables 
\be
x={\sqrt{2\mu} \cosh \sigma \over \sqrt{1-e^{2 \tau}}}
~,~~~~~~~~~~~~(1+e^{2t})(1-e^{2\tau}) = 1
\ee
brings the action for small fluctuations around this surface 
exactly to the static action in (\ref{action}).

The entire evolution of the Fermi surface is
described by a coordinate patch $\sigma \ge 0$ and 
$\tau \le 0$. The corresponding Penrose diagram is shown in
figure 1(b). 
Time dependence of the solution is now 
hidden in the presence of this boundary, since the 
effective action, (\ref{action}), does not depend on
on $\tau$ at all.  Even though nothing interesting happens
to the action at $\tau=0$, there is no reason to extend
pass the spacelike boundary, as the evolution of the original
system is fully captured by just this incomplete patch.  
Extending pass $\tau = 0$ has no meaning in the matrix model.

Another interesting solution we will encounter in the later
section is given by
\be
\label{opening}
(x-p) \left (x+p+e^{2t}(x-p)\right) ~=~ 2\mu~.
\ee
This describes a hyperbola which opens up to a straight line
(see figure 2(c)).
Again, the effective action can be brought into the static form
(\ref{action}), only this time the change of variables
is more involved,
\bear
t &<& 0, ~~
x= \pm {\sqrt{2\mu} \cosh \sigma \over \sqrt{1+e^{2 \tau}}}
~,~~~~~~~~~~~~(1-e^{2t})(1+e^{2\tau}) = 1~;
\\ 
t &>& 0,~~ x = ~~ {\sqrt{2\mu} \sinh \sigma \over \sqrt{e^{-2 \tau}-1}}
~,~~~~~~~~~~~~(e^{2t}-1)(e^{-2\tau}-1) = 1~.
\eear
The subtlety here is that we must include both sides of the
potential (allow both positive and negative $x$) as the solution
crosses $x=0$ at $t=0$.  The corresponding 
Penrose diagram can be seen in figure 
1(c). It consists of two pieces,
one for $t<0$ and one for $t>0$, joined by an identification
at the null boundaries.  We will have more to say about this in 
subsection 3.1.

The remarkable fact that the classical effective action is the same
for all above solutions suggests that perhaps
the equivalence holds at the quantum level as well.\footnote
{Notice that this property does not hold for all
possible solutions, but only for those which are given by conic sections 
\cite{Ernebjerg:2004ut}.}
This turns out to be true, and is one of the main points of this
paper. 

As a first step,
we shall introduce a bit of useful notation.  Consider two
coordinate systems, which we will refer to as A and B, linked by
the following transformation (given here together with its inverse for
later convenience)
\bear
&&\left \lbrace \begin{array}{rcl}
x_B &=& {1 \over \sqrt{1+e^{2t_A}}}~x_A \cr ~&~&~\cr
p_B &=&  \sqrt{1+e^{2t_A}}~p_A - {e^{2t_A}\over \sqrt{1+e^{2t_A}}} ~x_A
\end{array} \right .
\nn \\
&&\left \lbrace \begin{array}{rcl}
x_A &=& {1 \over \sqrt{1-e^{2t_B}}}~x_B \cr ~&~&~\cr
p_A &=&  \sqrt{1-e^{2t_B}}~p_B + {e^{2t_B}\over \sqrt{1-e^{2t_B}}} ~x_B
\end{array} \right .
\label{mapping} 
 \\
\nn \\
&&\left ( e^{2t_A} + 1 \right) \left ( 1-e^{2 t_B} \right) = 1 \nn
\eear
The coordinate transformation (\ref{mapping}) was chosen so that,
if
\be
{d \over d t_A} ~ x_A = p_A~~{\mathrm{and}}~~
{d \over d t_A} ~ p_A =  x_A~,
\ee
then
\be
{d \over d t_B} ~ x_B = p_B~~{\mathrm{and}}~~
{d \over d t_B} ~ p_B =  x_B~,
\ee
and therefore the Hamiltonian in both coordinates is just 
(\ref{hamiltonian}):
\be
H = \half p_A^2 - \half x_A^2 = \half p_B^2 - \half x_B^2~.
\label{hamiltonianAB}
\ee
As this transformation leaves the Dirac (or the Poisson) bracket
invariant
\be
[x_A, p_A] = [x_B,  p_B] = i~,
\ee
it can be treated as a change in either classical or
quantum phase space variables for the fermions.

It is easy to check that the transformation from B to A turns the
static solution in equation (\ref{static}) into the closing hyperbola
solution in equation (\ref{closing}), and that the inverse transformation
from A to B turns the static hyperbola into the opening hyperbola.
Actually, the latter is only true for $t_B < 0$; to obtain
the remainder of the evolution of the
opening hyperbola we need to analytically continue  (\ref{mapping}) 
in time.

This extended mapping from A to B,, valid for $t_B>0$ and $t_A<0$, is
\bear
&&\left \lbrace \begin{array}{rcl}
x_B &=& {1 \over \sqrt{e^{-2t_A}-1}}~x_A  \cr ~&~&~\cr
p_B &=&  -\sqrt{e^{-2t_A}-1}~p_A + {e^{-2t_A}\over \sqrt{e^{-2t_A}-1}} ~x_A
\end{array} \right .
\nn \\
&&\left \lbrace \begin{array}{rcl}
x_A &=& {1 \over \sqrt{e^{2t_B}-1}}~x_B \cr ~&~&~\cr
p_A &=& -\sqrt{e^{2t_B}-1}~p_B + {e^{2t_B}\over \sqrt{e^{2t_B}-1}} ~x_B
\end{array} \right .
 \\
\nn \\
&&\left ( e^{-2t_A} - 1 \right) \left ( e^{2 t_B}-1 \right) = 1 \nn
\label{mapping2} 
\eear

The above mapping takes a static hyperbola solution in A and turns it into
the second half of the evolution of the opening hyperbola in B,
when combined with a replacement of $\mu \rightarrow -\mu$.

Demanding that our mapping (\ref{mapping}) correctly connects the
static, opening and closing hyperbola solutions is not enough to
uniquely fix it.  For example, in \cite{Das:2007vfb}, a different
classical mapping was considered, based on the $W_\infty$ algebra acting 
on phase space.  What distinguishes (\ref{mapping}) from all other 
possible maps
is that the collective field $\eta(x)$ transforms trivially under it.
Therefore, (\ref{mapping})
preserves the form of the effective action
(\ref{action}).  We should  mention, however, that the mapping used in
\cite{Das:2007vfb} leads to the same quantum state as ours.

\section{Time dependent solutions in fermionic variables}

On the face of it, we have a map between two systems (either classical
or quantum) with the {\it same} Hamiltonian
which however evolve on a different time interval, since
$t_A$ runs from $-\infty$ to $\infty$ and $t_B$ runs from  $-\infty$
to $0$.  Using formulas in the Appendix, 
information contained in the mapping (\ref{mapping}) can
be summarized by a time dependent unitary operator which
transforms wavefunctions in the B system of coordinates into
those in the A system.  Using equation (\ref{app:U}), we see that
\bear
\label{Uxy}
U(t)&\equiv&
\exp \left (  i\ln \sqrt {1+e^{2t}} \left (x^2 - {xp+px \over 2} 
\right)\right) \\
&=&
\exp \left ( - i\ln \sqrt {1-e^{2\tau}} \left (x^2 - {xp+px \over 2} 
\right)\right)
\eear
does the job.   In order to avoid a large number of awkward
indices in the following discussion, we have set $t=t_A$ and
$\tau = t_B$.
$\tau(t) < 0$ is given by 
$\left ( e^{2t} + 1 \right) \left ( 1-e^{2 \tau} \right) = 1$.
To remind ourselves that $U$ evolves in time,
we will write is as either $U(t)$ or $U(\tau)$, whichever seems more
natural.  As $t$ and $\tau$ are linked a one-to-one function,
the choice of variable is cosmetic.

The fact that the two systems have the same
Hamiltonian is exhibited by the following 
nontrivial property of $U$:
\be
U(\tau) e^{-i(\tau-\tau_0)H}  = e^{-i(t-t_0) H} U(t_0)~,
\label{U-H}
\ee
where 
$\tau_0 \equiv \tau(t_0)$ (For example, if we take a
convenient choice of $t_0=0$ then $\tau_0 = -\ln\sqrt 2$.)
The above property of $U$ can be proven using a
special case of the Baker-Campbell-Hausdorff formula
(see Appendix for details).  
To put equation (\ref{U-H}) in words, 
a wavefunction can either
be first evolved from $\tau_0$ to $\tau$ and then 
acted upon with $U$ at time $\tau$, or be acted upon 
with $U$ at time $t_0$ and then evolved from
$t_0$ to $t$; the result will be the same.

To go from $A$ to $B$, we use $U^{-1} = U^\dagger$ 
\bear
U^\dagger(t) &\equiv&
\exp \left (  -i\ln \sqrt {1+e^{2t}} \left (x^2 - {xp+px \over 2} 
\right)\right) \\
&=&
\exp \left ( i\ln \sqrt {1-e^{2\tau}} \left (x^2 - {xp+px \over 2} 
\right)\right)~, \nn
\eear
and which has the property
\be
U^\dagger(t) e^{-i(t-t_0) H}  = e^{-i (\tau-\tau_0) H} U^\dagger(\tau_0)~.
\ee

\begin{figure}[t]
\parbox[l]{2in}{(a) 

\includegraphics[scale=0.25]{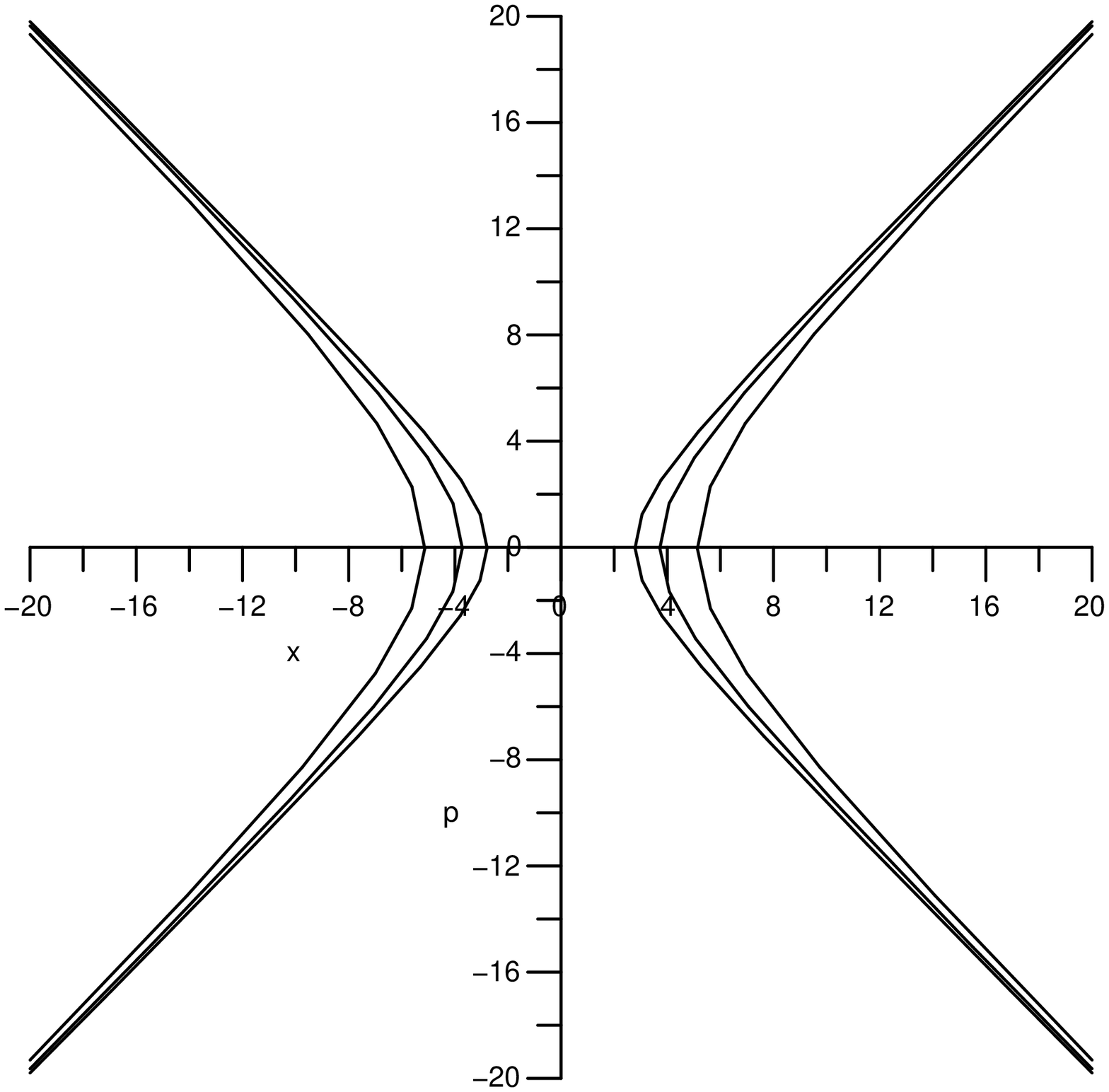}}
\parbox[l]{2in}{(b) 

\includegraphics[scale=0.25]{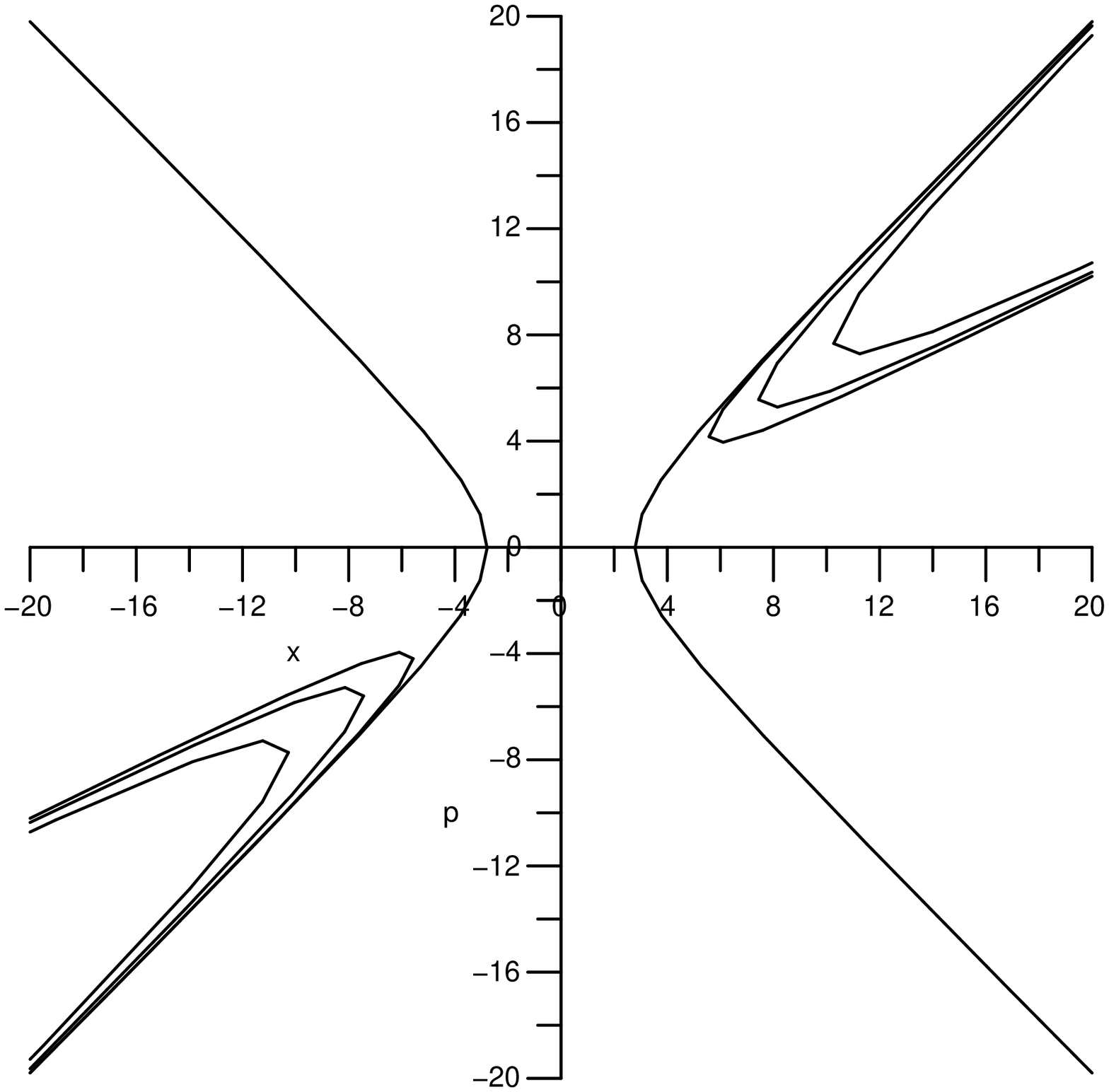}}
\parbox[l]{1.8in}{(c) 

\includegraphics[scale=0.25]{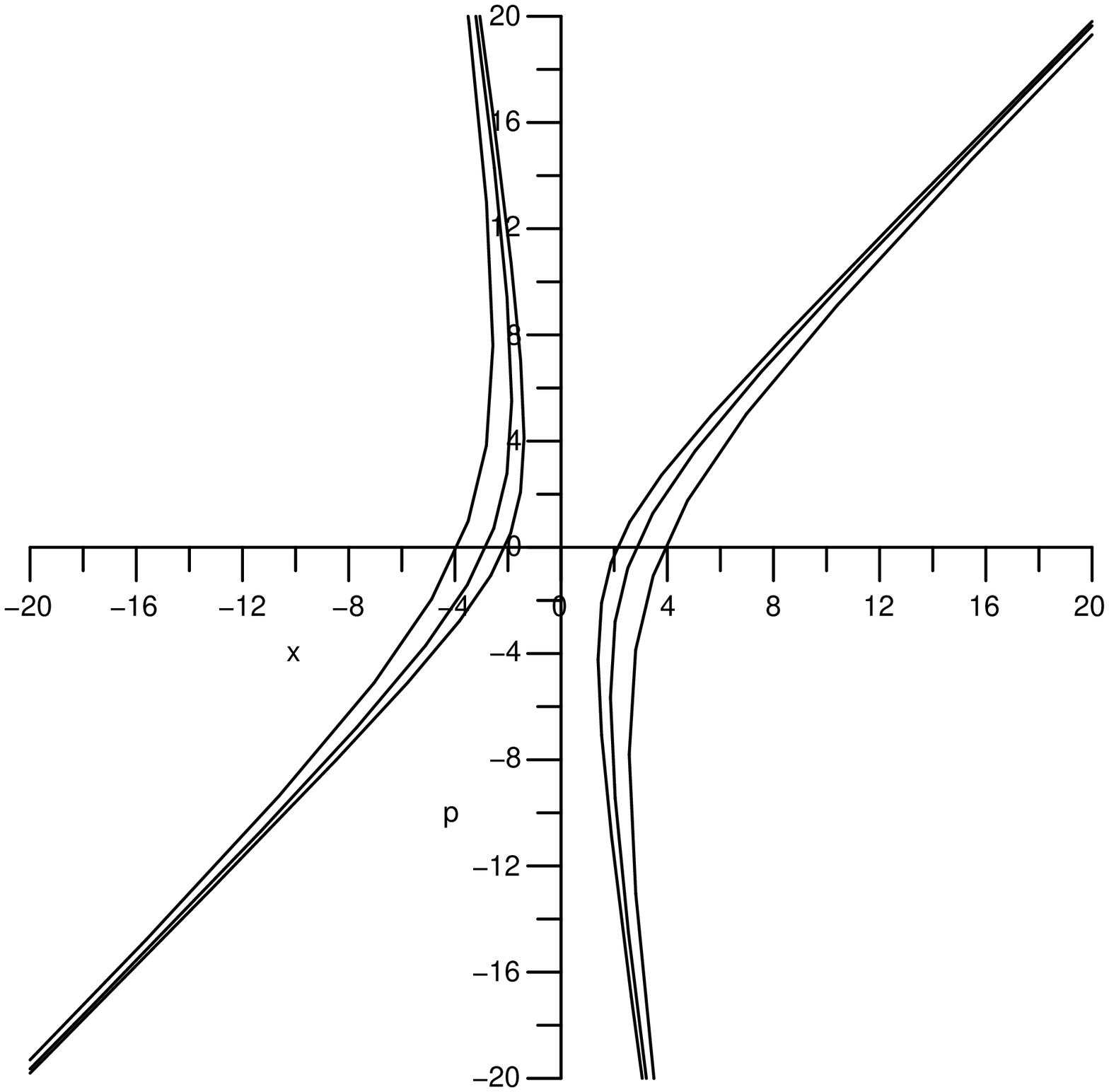}}

\caption{Rough contour plots of the absolute value of the
Wigner wavefunctions with $E=-10$.  In each case, the 
center hyperbola is where the Wigner wavefunction peaks
and the two other contours are at half hight.
(a) Static state $\psi_E(x)$;
(b) closing hyperbola state (\ref{wf:tdep}) at 
$\exp(2t)=3$, the additional hyperbola
shown is the half-hight contour from (a);
(c) closing hyperbola state (\ref{wf:tdep2}) at
$\exp(2\tau)=2/3$.
The data for the plots was computed from the definition of the Wigner
wavefunction in the chiral basis.}
\end{figure}

The unitary operator U allows us to write time
dependent wavefunctions corresponding to the
closing and opening hyperbola solutions, starting
from the well known static eigenfunctions.
Denote with $\psi_E(x)$ the eigenfunctions of the
Hamiltonian (\ref{hamiltonian}) with energy $E$:
\be
\label{diffeqn}
-{1 \over 2}(\partial_x^2 + x^2) \psi_E(x) = E \psi_E(x)~.
\ee
There are  two such eigenfunctions at each eigenenergy, 
even or odd under parity. Since the change of variables we consider in 
(\ref{mapping}) commutes with taking $x\rightarrow-x,~p\rightarrow-p$,
it is not necessary to worry about this degeneracy --- everything we say will
be true for the even and odd eigenfunctions separately.

Using again the formul\ae\ in the Appendix, we rewrite $U$ and
$U^\dagger$ in the following form
\be
U(\tau) \equiv (1-e^{2\tau})^{1/4} 
\exp \left (  {i\over 2} e^{2\tau} x^2\right)
\exp \left (  \ln \sqrt {1-e^{2\tau}} ~x {\partial \over \partial x}
 \right)~,
\ee
\be
U^\dagger(t) \equiv  (1+e^{2t})^{1/4} 
\exp \left (  -{i\over 2}e^{2t} x^2\right)
\exp \left (  \ln \sqrt {1+e^{2t}} ~x {\partial \over \partial x} \right)~.
\ee

This form makes the action of $U(t)$ on an arbitrary wavefunction
easy to read off.  When acting on the stationary wavefunction 
$\psi_E(x) e^{-iE\tau}$ with $U(t)$, we obtain
\be
\label{wf:tdep}
\Psi(x, t) \equiv U(t) \psi_E(x) e^{-iE\tau}
= (1+e^{2t})^{-1/4}
\exp\left({i \over 2}{e^{2t} \over 1 + e^{2t}}x^2 \right )
\psi_E\left({x \over \sqrt{1+e^{2t}}}\right) e^{-i E \tau(t)}~.
\ee
It is easy to check that (\ref{wf:tdep}) is a solution to the time dependent
Schrodinger equation with Hamiltonian (\ref{hamiltonian}), as 
long as we view $t$ as the appropriate time variable.
This wavefunction corresponds to the closing hyperbola, and
is valid for all $t$.

Similarly,
\bear
\label{wf:tdep2}
\tilde \Psi(x, \tau) &\equiv& U^\dagger(\tau) 
\psi_E(x) e^{-iEt} \nn \\
&=& (1-e^{2\tau})^{-1/4}
\exp\left(-{i \over 2}{e^{2\tau} \over 1 - e^{2\tau}}x^2 \right )
\psi_E\left({x \over \sqrt{1-e^{2\tau}}}\right) e^{-i E t(\tau)}~,
\eear
corresponds to the first half of the evolution of the opening hyperbola
(for $\tau<0$).
To obtain the second half of that evolution, we must analytically
continue in $\tau$, which will be done in subsection 3.1.

Let us investigate the form of the time dependent wavefunctions in some detail.

The static wavefunctions $\psi_E$s are known exactly \cite{Moore:1991sf},
but let us start with the large $x$ asymptotics.  
From the WKB approximation, at large $x$ the wavefunctions approach 
\be
\label{asympt}
\psi_E(x) \sim {1 \over \sqrt x} e^{\pm i x^2 / 2}
\ee
for all finite $E$.  Therefore
\be
\Psi(x) \sim {1 \over \sqrt x}
\exp\left({i \over 2}{e^{2t} \pm 1 \over  e^{2t}+1}x^2 \right )~.
\ee
For the upper sign, the asymptotic behaviour is the same as in
equation (\ref{asympt}), but for the lower sign, the behaviour
is markedly different.  This raises doubt about whether (\ref{wf:tdep})
can be written as a linear combination of $\psi_E(x)$s.  
If not, the time dependent wavefunctions $\Psi$ would be living in a different
Hilbert space from the $\psi_E$s, and our quantum equivalence
would be in trouble.  

Fortunately, this is not the right argument.  The question whether
these wavefunctions live in the same Hilbert space should be
answered by comparing the space of $L^2$ wavepackets that
can be build out of the `energy eigenbasis' in either case.
A moment of thought reveals that the Hilbert space is indeed the same.
Let's make it explicit.  Consider a wavepacket build out of
the static eigenfunctions $\psi_E(x)$, and denote it with $\varphi(x)$.
Now let's act on this wavepacket with the unitary operator $U(0)$,
(taking $t=0$ to be definite).  This gives us a new wavepacket
\be
\tilde \varphi(x) 
= (\mbox{phase})~(2)^{-1/4}
\exp\left(i x^2/4 \right )
\varphi\left({x \over \sqrt{2}}\right) ~.
\ee
There is no reason why this new wavepacket, also in $L^2$,
but formally in the Hilbert space of the closing hyperbola states, 
cannot be written 
as a linear combination of the static eigenfunctions $\psi_E(x)$.
We can calculate the Fourier coefficients as we always do,
and the integrals must converge, by the virtue of $\tilde
\varphi$ being in $L^2$.
The two Hilbert spaces are therefore the same.
Simply comparing asymptotic behaviour
was not enough because,  for any fixed $x$, there are
energies $E$ sufficiently negative that the asymptotics
do not apply.   The change of basis formula, which we will derive
in section 4, should also be interpreted in the sense of wavepackets.

\subsection{Analytic continuation through $\tau=0$}

This subsection is a detour away from the main line of the paper
and can be skipped.
 
 (\ref{wf:tdep}) 
describes a complete history of one fermion.
A collection of such fermions, one for each $E$ from minus infinity
up to some $\mu$ is the quantum state corresponding to the classical
solution (\ref{closing}).  We have obtained this wavefunction by a 
linear transformation from the static wavefunction, but in doing so,
we only used the evolution of the static state up to $\tau=0$.  What
about $\tau$ positive?  Formally, we can analytically continue the
change of variables in (\ref{mapping}) to positive $\tau$ by
replacing $t \rightarrow i\pi(2n+1)/2 - t$, where
$n$ is an arbitrary integer.
This will make the argument
in $\psi_E$ imaginary, so we need to understand
$\psi_E(ix)$.

Fortunately, $\psi_E(ix)$ is easy to deal with.
In the differential equation (\ref{diffeqn}), 
the variable $x$ can be thought of as a complex variable.
As long as we focus on either odd or even wavefunctions,
the solutions to (\ref{diffeqn}) are unique.  Substituting
$x \rightarrow ix$ in (\ref{diffeqn}) takes us back to the
same equation, but with $E \rightarrow -E$.  Therefore,
using uniqueness, we must have
$\psi_E(ix) \sim \psi_{-E}(x)$.  The magnitude of the
proportionality factor can be determined from the 
known behaviour of the properly normalized wavefunctions
at $x \ll \sqrt{|E|}$ \cite{Moore:1991sf} 
\bear
\psi_E(x) \sim {e^{-\pi E/2} \over E^{1/4}}~\cosh\left(
\sqrt{2E} ~x \right) ~~~&~~&\mathrm{for~even~wavefunctions,}
\\
\psi_E(x) \sim {e^{-\pi E/2} \over E^{1/4}}~\sinh\left(
\sqrt{2E}~ x \right)  ~~~&~~&\mathrm{for~odd~wavefunctions.}
\eear
Therefore,
\be
\psi_E(ix) = (\mathrm{phase}) ~e^{\pi E}~\psi_{-E}(x)~.
\ee
for any $\psi_E$ which is either even or odd under $x \rightarrow -x$.

With this result in hand, we can write, up to a constant, 
the wavefunction resulting from continuing $\tau$ through zero to
positive side in equation (\ref{wf:tdep})
\be
\label{wf:tdep:2}
\Psi(x, t) = (e^{-2t}-1)^{-1/4}
\exp\left({i \over 2}{e^{-2t} \over  e^{-2t}-1}x^2 \right )
\psi_{-E}\left({x \over \sqrt{e^{-2t}-1}}\right) e^{-i E \tau(t)}
\ee
where now $(e^{2\tau}-1)(e^{-2t}-1)=1$ 
and $t<0$. As can easily
be verified, this also is  a solution to the time dependent Schrodinger
equation.

The meaning of the analytic continuation through $\tau = 0$ becomes
more clear if we analyze the behaviour of the opening hyperbola
solution.  We will ignore again the overall normalization in the discussion,
as it has no bearing on the physics.

Equation (\ref{wf:tdep2}) gives the wavefunction corresponding to
the opening hyperbola, for the first half of its evolution, 
$\tau<0$.  We can analytically continue this
formula to positive $\tau$, where we obtain (up to an overall,
irrelevant normalization)
\be
\label{wf:tdep-open2}
\Psi(x, \tau) = (e^{2\tau}-1)^{-1/4}
\exp\left({i \over 2}{e^{2\tau} \over  e^{2\tau}-1}x^2 \right )
\psi_{-E}\left({x \over \sqrt{e^{2\tau}-1}}\right) 
e^{ i E t(\tau)}~,
\ee
which also satisfies the Schrodinger equation.  To check whether this
analytic continuation indeed gives the second half of the evolution of
the opening hyperbola, let's compare it with the wavefunction
obtained by transforming the stationary wavefunction with the
 the second mapping in section 2, equation (\ref{mapping2}).  Under
that transformation,
a stationary wavefunction in system A,  $\psi_E(x) e^{-Et}$, becomes
\be
\label{wf:tdep-open3}
\Psi(x, \tau) = (e^{2\tau}-1)^{-1/4}
\exp\left({i \over 2}{e^{2\tau} \over  e^{2\tau}-1}x^2 \right )
\psi_{E}\left({x \over \sqrt{e^{2\tau}-1}}\right) 
e^{- i E t(\tau)}~.
\ee
This is clearly the same wavefunction as (\ref{wf:tdep-open2}),
as long as we replace $E \rightarrow -E$, in agreement 
with the $\mu \rightarrow
-\mu$ replacement which is part of (\ref{mapping2}).

The meaning of the analytic continuation is now clear: if we are interested
in the evolution of the system B over the entire range of $t_B$, from
$-\infty$ to $+\infty$, we must continue past $t_A = +\infty$, or 
alternatively use a second mapping for the second half of the evolution 
(which is what was done in \cite{Das:2004aq}).  
These two approaches will lead to the
same answer.  Analytic continuation of 
$\tau \rightarrow  i \pi (2n +1)/2 - \tau$ 
is then the meaning we should assign 
to the identification of boundaries in the Penrose diagrams in figure 1(c)
(represented by an arrow there).

\section{Change of basis formula}

In this section, we will study  exact expressions for 
time dependent wavefunctions introduced in section 3, culminating in
an explicit formula giving the closing hyperbola wavefunction
as a linear combination of the static wavefunctions.
We will perform this analysis in the chiral
formalism, (first introduced in
\cite{Alexandrov:2002fh}), 
in which the form of the wavefunctions is simplest.  

Define $a_\pm$ to be
\be
a^\pm\equiv (x \pm p) / \sqrt 2~,
\ee
so that $[a^-, a^+] = i$ and $a_\mp = \pm i \partial / \partial_{a_\pm}$.
Our  mapping (\ref{mapping}) in these coordinates is
\bear
\label{mapping-apm}
a^-_B &=&  \sqrt {1 + e^{2t_A}} ~a^-_A \\
a^+_B &=&{1 \over \sqrt {1 + e^{2t_A}}} ~a^+_A -
{e^{2t} \over \sqrt {1 + e^{2t_A}}}~ a^-_A ~.
\eear 
The advantage of the chiral coordinates is that the Hamiltonian
is particularly simple
\be
H = \mp i \left( a^\pm {\partial \over \partial a^\pm} + \half \right)~,
\ee
and so are its eigenfunctions, 
\be
\psi_E(a^\pm) = a_\pm^{\pm iE - \half}~.
\label{chiral-wfn}
\ee
Including time evolution is also very simple. Any wavefunction of the
form $e^{\mp t/2}\varphi(e^{\mp t} a^\pm)$, for arbitrary $\varphi(\cdot)$,
is a solution to the time dependent Schrodinger equation.  In particular,
dressing up  energy eigenfunctions in (\ref{chiral-wfn}) with the 
proper time dependence gives
\be
\psi_E(a^\pm,t) = (a^\pm)^{\pm iE - \half} e^{-iEt} 
= e^{\mp t/2} (e^{\mp t} a^\pm)^{\pm iE - \half}
\label{wfn-tdep-a}
\ee
The unitary operator in equation (\ref{Uxy}), when written in terms of $a^\pm$,
is
\be
U(t) = \exp \left (
i \ln \sqrt {1+e^{2t}} 
\left ( a_-^2 + {a_+a_- + a_- a_+ \over 2}
\right )
 \right )~.
\ee
Using the formul\ae\ in the Appendix, we can rewrite this as
\be
U(t) = \left (1+e^{2t} \right )^{1/4}
\exp \left ( {i \over 2} e^{2t} a_-^2 \right )
\exp \left (\ln \sqrt {1+e^{2t}} 
 a_- {\partial \over \partial a_-}
 \right )~.
\ee

Acting with $U(t)$ on $\psi_E(a^-,\tau)$, we obtain the
wavefunction of the closing hyperbola in the $a_-$ basis:
\be
\Psi (a^-,t) =e^{t/2}
\exp \left ( {i \over 2} e^{2 t}(a^-)^2 \right )
  (e^t a^-)^{-iE - \half}~,
\label{wfn-closing-a}
\ee
where we have rearranged the wavefunction to exhibit an appropriate
form of time dependence.

In this basis, it turns out to be possible to figure out how
to express the wavefunction (\ref{wfn-closing-a}) as a linear
combination of wavefunctions of the form (\ref{wfn-tdep-a}).

Do accomplish this, we make use of the following identity:
\bear
e^{i z} &=& \int_C~ {ds \over 2\pi}~ z^{-is} ~e^{-\pi s / 2}~ \Gamma(is)
\\
&=& \lim_{A,B \rightarrow +\infty}
P.V. \int_{-A}^{B}~ 
{ds \over 2\pi}~ z^{-is} ~e^{-\pi s / 2}~ \Gamma(is)~+~\half~,
\label{identity}
\eear
Contour C runs along the real-s axis from $-\infty$ to $+\infty$
and below the pole at $s=0$.  This formula can
be obtained from the integral representation of the $\Gamma$-function
\cite[equation 8.312.2]{gr}, together with
  orthogonality conditions for the chiral wavefunctions 
(\ref{chiral-wfn}).  It holds for $z>0$, 
and has been verified numerically.

The convergence for $A,B \rightarrow  \infty$ is not
uniform in $z$.  The limit $B \rightarrow \infty$
can be taken in a uniform fashion, since the
integrand goes to zero rapidly for large positive $s$.
For $s$ large and negative, the integrand oscillates and
only goes to zero as $1/\sqrt{|s|}$. It is then necessary to 
restrict $A \gg z$.  
Without uniform convergence, 
we have to be careful when applying this
formula.

Using (\ref{identity}), we have that
\be
\exp\left ( {i \over 2} a^2\right ) a^{-\half - iE} =
\int_C {d\omega \over 4 \pi}~  2^{{i\over 2} (E-\omega)}
e^{{\pi \over 4}(E-\omega)}
\Gamma \left (-{i \over 2} (E-\omega) \right ) a^{-\half - i\omega}
\ee
and therefore
\be
\Psi (a^-,t) = U(t) \psi_E(a^-,\tau) =
\int d\omega  K(E-\omega) \psi_\omega(a^-,t)~,
\ee
where
\be
K(\nu) \equiv {1\over 4 \pi} ~ 2^{{i\nu\over 2} }
e^{{\pi \nu \over 4}} 
\Gamma \left (-{i\nu \over 2}  \right ) 
~+~ \half \delta(\nu)~.
\ee
At the end of section 3, we argued that  Hilbert spaces of the
closing hyperbola states and the static states are the same, since
the same $L^2$ wavepackets can be build in both cases.
The formula above should be read in that spirit: it links the expansions of
any given wavepacket in the two basis.  Focusing on wavepackets removes
any difficulty caused by the lack of uniform convergence.

For the sake of completeness, let's rewrite this result in the x-basis
\bear
\Psi(x, t) &=& (1+e^{2t})^{-1/4}
\exp\left({i \over 2}{e^{2t} \over 1 + e^{2t}}x^2 \right )
\psi_E\left({x \over \sqrt{1+e^{2t}}}\right) e^{-i E \tau(t)} 
\nn \\ &=&
\int d\omega K(E-\omega)
\psi_\omega(x,t)~.
\eear
The kernel $K$ is simply a representation of the unitary 
operator $U(t)$ in the appropriately time-evolving energy eigenbasis.

Notice that $K(\nu)$ decays exponentially for $\nu > 0$.
Therefore, energy eigenstates with energy greater than $E$ do not
enter into the closing hyperbola solution labeled by E.  This fact should
is illustrated in Figure 2(b): there, we can see that the
contours for the closing hyperbola state lie within ({\it{i.e.}}, at lower x)
the static hyperbola contour at the same E.  

\section{The Fermi sea and correlators}

We can now discuss the quantum state of the doubly scaled matrix model.
The fermionic field is defined as
\be
\Psi(x,t) = \sum_{E}~ \psi_E(x) ~e^{-iEt} ~c_E~,
\ee
where $c_E$ is an annihilation operator for a fermion with energy $E$,
$\lbrace  c_E, c_{E'}^\dagger \rbrace = \delta_{E,E'}$.  
Static ground state filled up
to the energy $\mu$ is defined as
\be
|\mu \rangle = \left ( \prod_{E < \mu}~ c_E^\dagger \right )
 |{\bf{0}}\rangle~,
\ee
where $|{\bf{0}}\rangle$ is the state with no fermions, 
$c_E |{\bf{0}}\rangle = 0$ for all $E$.  The operator
which creates a single fermion with a wavefunction $\varphi(x)$
at time $t$ is
\be
c_\varphi^\dagger =
\int dx~ \Psi^\dagger(x,t) ~\varphi(x) ~.
\ee
Therefore, the operator which creates a fermion in one of the
closing hyperbola states is given by
\be
c_{E,\mathrm{closing}}^\dagger \equiv 
\int dx~ \Psi^\dagger(x,t) ~\int d\omega K(E-\omega ) \psi_\omega(x,t)~
= \int d\omega K(E-\omega) c_\omega^\dagger~,
\ee
and the state corresponding to the closing hyperbola is
\be
|\mu,\mathrm{closing} \rangle = \left ( \prod_{E < \mu}~ 
\int d\omega K(E-\omega)~ c_\omega^\dagger \right )
 |{\bf{0}}\rangle~.
\label{state-in-H-of-MQM}
\ee
This formula shows that the decaying Fermi sea is a state in 
the Hilbert space of the matrix model, an important fact,
but  not useful for
computation of fermion correlators. 
More useful formul\ae\ can be obtained if we use
the linear transformation $U(t)$ instead of the kernel $K$.

To do that, let's define
\be
\tilde c_E = c_{E,\mathrm{closing}} \equiv 
\int dx~ \Psi(x,t) ~(U^\dagger(t) \bar \psi_E(x,\tau))~,
\ee
\be
\tilde c_E^\dagger = c_{E,\mathrm{closing}}^\dagger \equiv 
\int dx~ \Psi^\dagger (x,t) ~(U(t) \psi_E(x,\tau))~,
\ee
so that
\be
|\mu,\mathrm{closing} \rangle = \left ( \prod_{E < \mu}~ 
\tilde c_\omega^\dagger \right )
 |{\bf{0}}\rangle
\ee
and 
\bear
\Psi(x,t) = \sum_{E}~ (U(t)\psi_E(x) ~e^{-iE\tau}) ~\tilde c_E~.
\eear
Since $\lbrace\tilde c_E, \tilde c_{E'}\rbrace = \delta_{E,E'}$, 
any correlator of the form
\bear &&
A^{\textrm{closing}}
(x_1,t_1,\ldots,x_n,t_n;x_1',t_1',\ldots,x_n',t_n') = \\ \nn&&
\langle \mu,\mathrm{closing} |
\Psi^\dagger(x_1',t_1')\ldots
\Psi^\dagger(x_n',t_n')
\Psi(x_1,t_1)\ldots
\Psi(x_n,t_n)
|\mu,\mathrm{closing} \rangle 
\eear
can be computed as a corresponding correlator in the static state,
\bear &&
A^{\textrm{closing}}
(x_1,t_1,\ldots,x_n,t_n;x_1',t_1',\ldots,x_n',t_n') = \\ \nn&&
\langle \mu|
(U^\dagger(t_1') \Psi^\dagger(x_1',\tau_1'))\ldots
(U^\dagger(t_n') \Psi^\dagger(x_n',\tau_n'))
(U(t_1) \Psi(x_1,\tau_1))\ldots
(U(t_n) \Psi(x_n,\tau_n))
|\mu \rangle ~.
\eear
We have seen that
\be
U(t)\Psi(x,\tau) = \left (1+e^{2t} \right)^{-1/4}
\exp \left ( {i \over 2} {e^{2t} \over 1 + e^{2t}} x^2 \right )
\Psi\left ({x \over \sqrt {1+e^{2t}}}, \tau(t) \right )
\ee
and therefore
\bear && 
A^{\textrm{closing}}
(x_1,t_1,\ldots,x_n,t_n;x_1',t_1',\ldots,x_n',t_n') = \nn \\ \nn&&
\prod_{k=1}^n \left (1+e^{2t_k} \right)^{-1/4}
\prod_{k=1}^n \left (1+e^{2t_k'} \right)^{-1/4}
\prod_{k=1}^n 
\exp \left ( {i \over 2} {e^{2t_k} \over 1 + e^{2t_k}} x_n^2 \right )
\prod_{k=1}^n 
\exp \left (- {i \over 2} {e^{2t_k'} \over 1 + e^{2t_k'}}(x_n')^2 \right )
\times \nn \\ &&
A^{\textrm{static}}
\left (
{x_1 \over \sqrt {1+e^{2t_1}}}, \tau_1,
\ldots,
{x_n \over \sqrt {1+e^{2t_n}}}, \tau_n;
{x_1' \over \sqrt {1+e^{2t_1'}}}, \tau_1',
\ldots,
{x_n' \over \sqrt {1+e^{2t_n'}}}, \tau_n' \right )~.
\eear
Correlators in the static background are well known, see
for example \cite{Moore:1991sf}.

This  formula is one of the main results of this paper, and is
a generalization of the formul\ae\ in \cite{Das:2007vfb}.

As a test, and a demonstration of this result, let us now compute
the equal time correlator
\bear
&& A^{\textrm{closing}}(x,t;y,t)= 
\langle \mu,\mathrm{closing} |~
\Psi^\dagger(y,t)\Psi(x,t)~
|\mu,\mathrm{closing} \rangle = \\ \nn  &=&
\left (1+e^{2t} \right)^{-1/2}
\exp \left ( {i \over 2} {e^{2t} \over 1 + e^{2t}} (x^2-y^2) \right )
A^{\textrm{static}}
\left (
{x \over \sqrt {1+e^{2t}}}, \tau;
{y \over \sqrt {1+e^{2t}}}, \tau \right)
\eear
This correlator is related to density of  fermion eigenvalues in the
x--p plane via a well known formula for the expectation
value of the Wigner operator in the context of the c=1 models
(see \cite{Dhar:1992rs} and references therein)
\be
\rho(x,p,t) = \int dy ~{e^{-iyp} \over 2\pi}~ \langle
\Psi^\dagger(x+y/2,t)
\Psi(x-y/2,t) \rangle~.
\ee
After a short calculation, we conclude that
\be
\rho^{\textrm{closing}}(x,p,t) = 
\rho^{\textrm{static}}\left (
{x\over\sqrt{1+e^{2t}}},~
\sqrt{1+e^{2t}} p - {e^{2t}\over\sqrt{1+e^{2t}}} x~
,\tau\right)~.
\label{density closing}
\ee
Taking the classical approximation where the density for
a static hyperbola is simply
\be
\rho^{\textrm{static}}_{\mu}(x,p,t) = \left \lbrace \begin{array}{ll} 
1~~~ & \textrm{for}~ x^2 - p^2 > \mu \\
0~~~ & \textrm{otherwise.}
\end{array} \right .
\ee
the above equation gives
\be
\rho^{\textrm{closing}}(x,p,t) = 
\left \lbrace \begin{array}{ll} 
1~~~ & \textrm{for}~ (x - p)(x+p - e^{2t}(x-p)) > \mu \\
0~~~ & \textrm{otherwise.}
\end{array} \right .
\ee
which is the same answer we would have obtained if we simply used 
the classical transformation (\ref{mapping}) on $\rho^{\textrm{static}}$.

A formula analogous to (\ref{density closing}) can be derived for
products of the Wigner operator, relating their correlators in
the closing hyperbola state to the correlators in the static
state via the classical mapping (\ref{mapping}).

\section{Discussion and extensions}

The same quantum evolution has been presented
in this paper in several different ways, which leads 
to the following ambiguity.  Let's say someone
presents us with a stationary wavefunction in the upside down
harmonic oscillator potential.  Without any further information,
it is not clear whether this wavefunction is meant to describe 
simply the stationary state, or the closing hyperbola state (in which case
we should interpret time as ending at zero), or the opening
hyperbola state (in which case we should analytically continue
the evolution of the system pass the time = infinity mark,
as can be seen in Figure 1(c) and was discussed in
section 3.1).

There is hope that gravity resolves this ambiguity.
After all, before it can describe string theory, the matrix model
must be augmented by a leg-pole transform, which encodes
gravitational and other interactions 
\cite{Polchinski:1994mb,Klebanov:1991qa}.  
Our analysis does not  capture everything about 
time dependent solutions to gravitational effective action.
Only once the time dependent Fermi sea profile is translated into a valid
background for dilaton gravity (and string theory), can 
additional information, such as 
the conformal factor for the metric, and the
behaviour of the dilaton, resolve this ambiguity.  
Unfortunately, such an analysis is
beyond the scope of this note.  We have taken a first 
necessary step towards it, by expressing the
closing hyperbola solution as a state in the Hilbert space of
the matrix model.

With our explicit formula (\ref{state-in-H-of-MQM}), it should be possible,
at least in principle, to bosonize the closing hyperbola 
quantum state, and to obtain a quantum state in the
bosonic collective theory which is closely related to
the string theory tachyon.  It might even be possible
to find the string theory background which corresponds to this solution.
The main obstacle is the currently incomplete understanding of the
leg-pole transform linking the collective
field to the tachyon.

\begin{center}---\end{center}

To keep the algebra simple, we have considered here only those
time dependent solutions of the matrix model which approach
the static solution in the infinite past.  As a result, 
of the two quantum mechanical descriptions under consideration,
one had a time variable running over the entire real line,
and the other had semi-infinite time.  It is possible to generalize
the discussion in the paper to a more involved situation in which
the time dependent solution has time reversal symmetry, and thus
diverges from the static one in both its
past and its future.  Then, one of the quantum mechanical
descriptions has compact time: the time interval over 
which the evolution happens is finite.

In \cite{Das:2007vfb}, 
an entire family of opening and closing hyperbola solutions
was discussed.  For each member of this family there exists a 
unitary operator $U$ which translates between time
evolution of the original matrix model and a new quantum system where
the time dependent solution  appears static.  The
quantum equivalence between two system discussed in the present paper
can be generalized this way to an entire family of equivalences.

One might wonder whether the results could be extended even further
than that.  What if we tried to treat more general Fermi surfaces,
obtained by acting with higher order operators in the $W_\infty$
algebra?  The effective action in that case could not be brought
into a static form  \cite{Ernebjerg:2004ut}, and therefore 
any equivalence would have to be between systems
with different Hamiltonians.  It would nonetheless be interesting to
investigate such a possibility.
Another interesting extension would be to consider the 
droplet solution  \cite{Ernebjerg:2004ut}.  

\addcontentsline{toc}{section}{Appendix}

\section*{Appendix}

Here we gather, for reference, a number useful formul\ae\ 
which are used throughout the paper.

\addcontentsline{toc}{subsection}{Baker-Campbell-Hausdorff formula}
\subsection*{Baker-Campbell-Hausdorff formula}
This formula states that for two elements $X$ and $Y$ of any algebra,
we can write $\exp(X) \exp(Y)$ as $\exp(X + Y + ~\ldots)$, where  the
$~\ldots~$ is build entirely out of nested commutators of $X$ and $Y$.
The most widely used version of this formula is 
\be
[X,Y] = \theta 
~~~~ \Rightarrow ~~~~
\exp(X) \exp(Y) = \exp\left(X + Y +
{\theta \over 2} \right )~,
\ee
applicable if $\theta$ is a c-number (in the center of the algebra).

We are going to need a little more.  The following formul\ae\
can be derived explicitly, for example in the sl(2) algebra,
\bear
[X,Y] = sY
&~~~~ \Rightarrow ~~~~&
\exp X \exp Y = \exp
\left (X + {s \over 1- e^{-s} }~ Y \right )~, \nn \\
&\mbox{and}& 
\exp Y \exp X = \exp
\left (X + {s \over  e^{s} -1 }~ Y \right )~.
\label{BCH1}
\eear
Using the  above two formul\ae\, we can also show that
\bear
&&[X,Y] = s(X+Y)
~~~~ \Rightarrow ~~~~ \nn \\ &&
\exp(\alpha X) \exp(\beta Y) = \exp \left [
(\alpha-\beta) \left (
{e^{s\alpha} - 1\over e^{s(\alpha-\beta)}-1  }~X~ +~
{e^{s\beta} - 1\over 1- e^{s(\beta-\alpha)}  }~Y 
\right )\right ]~,
\label{BCH2}
\eear
which is used to prove (\ref{U-H}).

\addcontentsline{toc}{subsection}{Change of canonical variables}
\subsection*{Change of canonical variables}

Let $u$ and $v$ be two canonical variables with $[\hat u,\hat v] = i$.
(We will use hats on operators here to make things more clear.)
We want to see how a wavefunction $\psi(u) \equiv \langle u | \psi \rangle$
corresponding to a state $|\psi\rangle$ is related to
 $\Psi(U) \equiv \langle U | \psi \rangle$.  We will assume that 
$\hat U,\hat V$ are
related to $\hat u,\hat v$ by
\be
\left ( \begin{array}{c}
\hat U \cr \hat V
\end{array} \right ) =
\left ( \begin{array}{cc}
a & b\cr c & d
\end{array} \right )
\left ( \begin{array}{c}
\hat u \cr \hat v
\end{array} \right ) ~.
\ee
We will take $ad - cb = 1$ so that $[\hat U, \hat V] = i$, 
and assume that $a$ is positive.

The relationship between $\psi(u)$ and $\Psi(U)$ depends on whether
$b$ is zero or not.  Let's start with the simpler case of $b=0$.  We 
then have $d = 1/a$ and $c$ is arbitrary.  It then follows that
\be
\Psi(U) = {1 \over \sqrt{a} }e^{{i \over 2}c d U^2} \psi(U/a)~.
\label{wfn-formula}
\ee
One way to obtain this formula is to define a unitary operator
\be
\exp(D) \equiv 
\exp\left( - i {\ln(a) \over 2}~  
\left ( \hat u \hat v + \hat v  \hat u \right ) 
- i {ac\ln(a) \over 1-a^2} \hat u^2\right ) 
\label{app:U}
\ee
which has the property that
\be
%& \exp \left (  i\ln(a)  (\hat u \hat v + \hat v \hat u ) +
% i  {ac\ln(a) \over 1-a^2} \hat u^2 \right )
%{\hat u \choose \hat v}
%\exp \left ( - i\ln(a)  (\hat u \hat v + \hat v \hat u ) -
% i  {a c\ln(a) \over 1-a^2}\hat u^2 \right ) & \nn \\
%& & \nn \\ &= 
\exp(-D) {\hat u \choose \hat v}  \exp(D) =
\left ( \begin{array}{cc}
a & 0\cr c & d
\end{array} \right )
\left ( \begin{array}{c}
\hat u \cr \hat v
\end{array} \right ) ~. 
\ee
We now make use of the one of special cases of the 
Baker-Campbell-Hausdorff formula, (\ref{BCH1})  to obtain
\bear
\exp(D) &=& 
\exp \left ({i \over 2}  c d u^2\right ) 
\exp\left( - {\ln(a) \over 2}~  
\left ( u {\partial \over \partial u}
+ {\partial \over \partial u}  u \right ) \right ) 
 \nn \\
&=& {1 \over \sqrt a}
\exp \left ({i \over 2}  c d u^2\right ) 
\exp\left( - \ln(a)~  
u {\partial \over \partial u} \right ) 
  ~,
\label{app:U:2}
\eear
from which equation (\ref{wfn-formula}) can be read off easily.

For completeness, let us remark that
for $b \neq 0$, we have
\be
\langle U | u \rangle = {1 \over \sqrt{2\pi b}} e^{ {i \over 2 b} 
(a u^2 - 2 u U + d U^2)}~,
\ee
and the relationship between $\psi$ and $\Psi$ is given by
\be
\Psi(U) = \int du  ~ \langle U | u \rangle ~\psi(u)~.
\label{app1}
\ee
If needed, this formula can be used to relate the wavefunctions in the 
chiral basis to those in the position basis.

%%%%%%%%%%%%%%%%%%%%%%%%%%%%%%%%%%%%%%%%%%%%%%%%%%%%%%%%%%%%%%%%%%%%
%  A C K N O W L E D G M E N T S                                   %
%%%%%%%%%%%%%%%%%%%%%%%%%%%%%%%%%%%%%%%%%%%%%%%%%%%%%%%%%%%%%%%%%%%%

\section*{Acknowledgments}
The author would like to acknowledge helpful conversations with
Sumit Das, Lior Silberman, Moshe Rozali and Don Witt.  This work
was supported by the Natural Sciences and Engineering Research
Council of Canada.

%%%%%%%%%%%%%%%%%%%%%%%%%%%%%%%%%%%%%%%%%%%%%%%%%%%%%%%%%%%%%%%%%%%%
%  B I B L I O G R A P H Y                                         %
%%%%%%%%%%%%%%%%%%%%%%%%%%%%%%%%%%%%%%%%%%%%%%%%%%%%%%%%%%%%%%%%%%%%

\bibliographystyle{JHEP}

\bibliography{my}

\end{document}